\documentclass[a4paper,12pt]{article}
\pdfoutput=1
\usepackage{amssymb,amsfonts}
\usepackage{graphicx}
\usepackage{multirow}
\usepackage[english]{babel}
\usepackage{amsmath}
\usepackage{graphicx}
\usepackage{color}
\usepackage{amssymb}
\usepackage{hyperref}
\usepackage{dcolumn}% Align table columns on decimal point
\usepackage{bm}% bold math
\usepackage{slashed}
\usepackage{url}
\usepackage{epsfig}
\usepackage{amsmath}
\usepackage{amssymb}
\usepackage{amsfonts}
\usepackage{bbm}
\usepackage[footnotesize]{caption}
\usepackage{mathrsfs}
\usepackage[font=scriptsize]{subcaption}
\usepackage{color}
\usepackage{comment}
\usepackage{mathtools}
\usepackage{braket}
\usepackage{cite}
\usepackage{hyperref}
\usepackage{multirow}
\usepackage{relsize}
\usepackage{fullpage}
\usepackage{makecell}

\begin{document}

\begin{center}
{\bf\Large  Left Right Model from Gauge Higgs Unification with Dark Matter} \\[12mm]
Francisco~J.~de~Anda
\footnote{E-mail: \texttt{franciscojosedea@gmail.com}}\\[-2mm]
\end{center}

\vspace*{0.2cm}
\centerline{$^{\dagger}$ \it
Departamento de F{\'i}sica, CUCEI, Universidad de Guadalajara, M{\'e}xico}
\vspace*{1.20cm}

\begin{abstract}
We propose a five dimensional model based on the idea of Gauge Higgs Unification with the gauge group $SO(5)\times U(1)$ in Randall-Sundrum spacetime.  We obtain a Left-Right symmetric model with a stable scalar indentified as a dark matter candidate. This stable scalar obtains a vacuum expectation value that gives mass to fermions in the bulk through the Hosotani Mechanism. There is a scalar localized on a brane and gives contributions to fermion masses. This scalar fits the observed Higgs boson data. We are able to fit all the Standard Model observables while evading constraints.
\end{abstract}

\section{Introduction}

In the Standard Model (SM), the largest sources of free parameters are the set of Higgs to fermions couplings and the Higgs potential. We would like to understand how these parameters appear from a more fundamental theory. 

One proposal for that is Gauge Higgs Unification (GHU), where one begins with an extra dimensional theory and the Higgs boson is the higher dimensional component of a gauge field \cite{Sakamura:2007qz,Medina:2007hz,Hall:2001zb,Gogoladze:2003bb,Hosotani:2006qp,Hosotani:2008je,Kojima:2011ad,Scrucca:2003ut,Antoniadis:2001cv}. In that case, the fermion-boson couplings as well as the scalar potential would be fixed by the gauge symmetry.

The most popular models begin with a gauge symmetry based on the group $SU(3)$ or $SO(5)$. These groups are able to generate a scalar with Higgs-like transformation properties under the SM group, but these groups can not accomodate its observables. To generate a realistic model one needs to make use of extra symmetries and fields \cite{Panico:2005dh}. One of such models is the so called Hosotani-Oda-Ohnuma-Sakamura (HOOS) model which is able to fit the measurements of the Higgs boson but requires many exotic fermions to do so \cite{Hosotani:2008tx, Hosotani:2012qp}.

The HOOS model has many interesting features which make it appealing. It has left-right (LR) symmetry that complies with electroweak precision tests. The scalar arising from the GHU idea, which we will call Hosotani scalar, is stable if one only has the SM fermionic content. The HOOS model adds many exotic fermions to fit the observed Higgs boson and make it unstable.

The Higgs boson is not easily obtained from GHU. The scalar obtained from the GHU idea is not the Higgs boson but another scalar field with special properties that may solve other problems, like dark matter \cite{Panico:2008bx}.

We work with a slight modification of the HOOS model. We obtain a LR model from this model \cite{Chacko:2005un,Mohapatra:1974gc,Mohapatra:1974hk,Senjanovic:1975rk} where the Hosotani scalar is not the Higgs boson but one of the scalars that break the LR symmetry. This scalar is stable and a dark matter candidate. We add two more scalars on the brane which contribute to LR symmetry breaking, the Higgs boson comes from one of them. This Higgs-like boson has modified Higgs to gauge boson couplings, with respect to the SM.

All the SM fermions live in the bulk and obtain mass contributions from the Higgs boson and the Hosotani scalar.

\section{$SO(5)\times U(1)$ model in the bulk}

The model is set up on the Randall - Sundrum (rs) warped spacetime with metric \cite{Randall:1999ee}:
\begin{equation}
ds^2 = e^{−2\sigma(y)} \eta^{\mu\nu}dx_\mu dx_\nu + dy^2,
\end{equation} 
where $\sigma(y)=\sigma(y+2L)$ and $\sigma(y)=k|y|$ for $|y| \leq L$.

Te fundamental region for the fifth dimension is $0\leq y\leq L$ and is limited by the UV brane at $y=0$ and the IR brane at $y=L$.

The model is based on the gauge group $SO(5) \times U(1)$ and the corresponding gauge fields propagate in the bulk. 

After compactifying the fifth dimension we obtain the Kaluza-Klein (KK) mass scale \cite{Hosotani:2008tx}:
\begin{equation}
m_{KK}=\frac{\pi k}{e^{kL}-1}\simeq \pi k e^{-kL}
\end{equation}
that must comply with extra - dimensional constraints coming from electroweak precision tests \cite{Beringer:1900zz}.

There are three 5D fermions $\Psi_{u,d,e}$ that live in the bulk. They transform as 
\begin{equation}
\Psi_u\sim[\textbf{5}, 2/3] ,~~\Psi_u\sim[\textbf{5}, -1/3],~~\Psi_e\sim[\textbf{5}, -1] \label{femo}
\end{equation}
under $SO(5)\times U(1)$. These fermions generate the scalar potential through the Hosotani mechanism \cite{Sakamura:2007qz,Carena:2007ua}.

The action for this model is:
\begin{equation}
S=\int \sqrt{g}d^5 x \Bigg(\sum_{k=u,d,e}(i\bar\Psi_k\slashed D \Psi_k-c_k\sigma'\bar\Psi_k \Psi_k)-\frac{1}{4}Tr[F^{MN}F_{MN}]-\frac{1}{4}B^{MN}B_{MN}\Bigg),
\end{equation}
where $F^{MN}$ is the field strength of $SO(5)$, $B^{MN}$ is the field strength of $U(1)$ and $c$ is the kink mass of the fermion in the bulk.

\subsection{Orbifolding, Hosotani phases and H parity}

When we compactify, the components $A_5$ of the gauge fields $A_M\in so(5)\times u(1)$ behave as scalars. We are free to choose the boundary conditions:
\begin{eqnarray}
\nonumber A_\mu(x,y_i-y) &=& P_i A_\mu(x,y_i+y) P_i^{-1}~,
\\ A_5(x,y_i-y)&=&-P_i A_5(x,y_i+y) P_i^{-1}~,
\\ \nonumber \Psi(x,y_i-y)&=&\eta^i \gamma^5 P_i\psi~,
\end{eqnarray}
where $P_i$ are two $5\times 5$ matrices, one for each brane, that obey $P_i^2, \eta^i=\pm 1$. The gauge symmetry on the branes will be reduced to the gauge transformations which satisfy:
 \begin{equation} P_i = \Omega(x,y_i-y) \, P_i \, \Omega^\dagger (x,y_i+y) ,\label{orsim}\end{equation}
where $\Omega$ is a gauge transformation from $SO(5)\times U(1)$.

We choose $P_i=diag(1,1,1,1,-1)$ and $\eta^i=1$. These conditions break $SO(5)\times U(1)\to SO(4)\times U(1)$ on the branes. The four broken generators now behave as real scalars that transform as a $[\textbf{4},0]$ under the remaining gauge group.

After breaking $SO(5)$ by orbifolding, the broken generators may obtain a vacuum expectation value (VEV) through the effective potential \cite{Oda:2004rm,Hosotani:2005fk}. We may align the VEV with a specific generator $\lambda^H$ so the scalar that obtains a VEV can be written as:
\begin{equation}
A_5^H(x,y)=\theta_H(x)h(y)\lambda^H,
\end{equation}
where $\theta_H(x)$ is a dimensionless scalar and $h(y)$ is its fifth dimensional profile, normalized as $\int_0^L h(y)=1$.

Now let's study the transformation
\begin{equation}
\Omega(y)=e^{i\alpha \int_0^y h(y) \lambda^H}
\end{equation}
that belongs to the broken generator of the group and changes $\theta_H\to\theta_H-\alpha$.
We see that for specific values of $\alpha$ this may be a symmetry of the model if it satifies Eq.(\ref{orsim}). In general this will be a symmetry for $\alpha=2\pi$, hence the name of Hosotani Phase for $\theta_H$. Since we are working only with tensorial representations of $SO(N)$, it is enough for $\alpha=\pi$ to be a symmetry of the system.

When we obtain the effective lagrangian for this scalar, namely \cite{Hosotani:2007qw}:
\begin{equation}
{\cal L}_{eff}  = - V_{eff} ( \theta_H) 
                 - m_W^2( \theta_H) W^\dagger_\mu W^\mu
                 - \frac{1}{2} m_Z^2( \theta_H) Z_\mu Z^\mu 
                 - \sum_{a,b} m^F_{ab}( \theta_H) \bar{\psi}_a  \psi_b ~.
\end{equation}
The functions of $\theta_H$ have the symmetries:
\begin{eqnarray}
V_{eff} ( \theta_H + \pi) = V_{eff} ( \theta_H) 
 = V_{eff} (-  \theta_H) , \nonumber
\\
m_{W, Z}^2( \theta_H + \pi) = m_{W, Z}^2 ( \theta_H) 
= m_{W, Z}^2 (-  \theta_H),\\
\nonumber 
m^F_{ab}( \theta_H + \pi) = - m^F_{ab} ( \theta_H) 
= m^F_{ab}  (-  \theta_H) ,
\label{effV1}
\end{eqnarray}
due to the symmetry of the extra dimension and the fact that $\theta_H$ is a phase.

In this model, with massive fermions in the bulk and the chosen gauge symmetry the VEV happens to be $\langle \theta_H \rangle=\pi/2$ \cite{Haba:2008dr, Hosotani:2008tx}. 

If we write $\theta_H=\frac{\pi}{2}+H_h/f$ where $H_h$ is the scalar field and $f$ is a dimensionful constant,
the above mentioned functions of the effective lagrangian behave like:
\begin{equation}
F\left(\frac{\pi}{2}+H_h/f\right)=F\left(\frac{\pi}{2}-H_h/f\right),
\end{equation}
therefore they do not contain odd powers of $H_h$. This is called H-parity ($P_H$). This is an enhanced symmetry that remains after ElectroWeak Symmetry Breaking (EWSB) $SU(2)_W\times U(1)_Y\to U(1)_{EM}\times P_H$.

Since $H_h$ is odd under H-parity and every other field is even, this scalar field is stable \cite{Hosotani:2010rg, Hosotani:2010hx}.

\section{Scalars in the model}

The IR brane has gauge symmetry $SO(4)\times U(1) \simeq SU(2)_R\times SU(2)_L \times U(1)_X$. Due to the rs metric, the dimensioful parameters are naturally small in this brane, solving the hierarchy problem. 

 The broken generators behave as a scalar that transforms under the remaining gauge group as:
\begin{equation}
H_h\sim[\textbf{2},\textbf{2},0]~.
\end{equation}
To break the LR symmetry completely we need more scalars. We add two scalar fields on the IR brane, transforming like:
\begin{eqnarray}
H_R\sim[\textbf{2},\textbf{1},1/2]~,
\\ \nonumber H_L\sim[\textbf{2},\textbf{2},0]~.
\end{eqnarray}
Each of these scalars has its usual symmetry breaking potential that gives them VEVs $v_{L,R}$, breaking $SU(2)_{L,R}\times U(1)_X$ respectively.
 
There are stringent constraints on the masses for the $W_R$ forcing $v_R\gtrsim 8$ TeV \cite{Beringer:1900zz, Maiezza:2010ic, Bambhaniya:2013wza}. The left scalar obtains a VEV $v_L\sim O(100GeV)$ so there is a two order of magnitude difference between these two VEVs. This must be tuned in the model.

Since we work in the low energy effective theory, where the KK modes are integrated out, the $H_R$ scalar and the $W_R^\pm, Z_R$ gauge bosons should be integrated out too, leaving us with the symmetry $SU(2)_L\times U(1)_Y$ on the IR brane with $Q_Y=T^3_R+Q_X$. The scalars transform, under the remaining gauge group, as:
\begin{equation}
H_h\sim H_L\sim [\textbf{2},1/2]
\end{equation}
and effectivelly we have a Two Higgs Doublet Model (THDM).

The W boson receives its mass from the VEV of the Hosotani scalar in the bulk and the left scalar VEV on the brane $(v_L)$ so it looks like \cite{Medina:2007hz,Haba:2008dr,Hosotani:2008tx}:
\begin{equation}
S=\int \sqrt{g} d^5 x w^+(y)w^-(y)\Big(\frac{m_{KK}}{\pi L \sqrt{kL} }+\delta(y-L)\frac{gv_L}{2}\Big)W^+(x)W^-(x),
\end{equation}
where the $w(y)$ are the gauge boson's profiles in the fifth dimension. This means that the mass term that comes from the brane will have additional input from the profiles after normalizing the four-dimensional fields. Luckily the profiles of the W and Z in the fifth dimension are practically constant and there is not much input from them to the mass parameter nor to the charged currents \cite{Hosotani:2007qw} . 

The VEV coming from the Hosotani mechanism may be aligned so that the mixing terms between $W_R,Z_R$ and $W_L,Z_L$ are practically zero \cite{Hosotani:2008tx,Senjanovic:1975rk}. 

The masses for the W and Z are given just like in a THDM with $\rho=1$ and the Weinberg angle has its value fixing the coupling constant of $U(1)_X$.

Due to H-parity there are no mixing terms between scalars. Since the $H_h$ scalar breaks the symmetry in the bulk, it contains the Goldstone bosons so we work with it as:
\begin{equation}
H_h=\left(\begin{array}{c}0\\\frac{v\frac{\pi}2+H(x)}{\sqrt{2}}\end{array}\right), \hspace{10mm}H_L=\left(\begin{array}{c}H^+\\\frac{v_L+h(x)+iA_0(x)}{\sqrt{2}}\end{array}\right),
\end{equation}
where the extra $\pi/2$ comes from the Hosotani Mechanism and $H(x)$ is the remaining neutral scalar field in $H_h$ (to be called Hosotani scalar too for the remaining of the paper).

Compactifying the extra dimension breaks 5D vectors into 4D vectors and scalars.  After intergrating out heavy particles (KK modes and right $SU(2)_R$ gauge bosons) we obtain effective interaction terms between the remaining scalars. This can be seen in processes show in figure \ref{fig:fe}:

\begin{figure}[!h]
\centering
 \includegraphics[scale=.5]{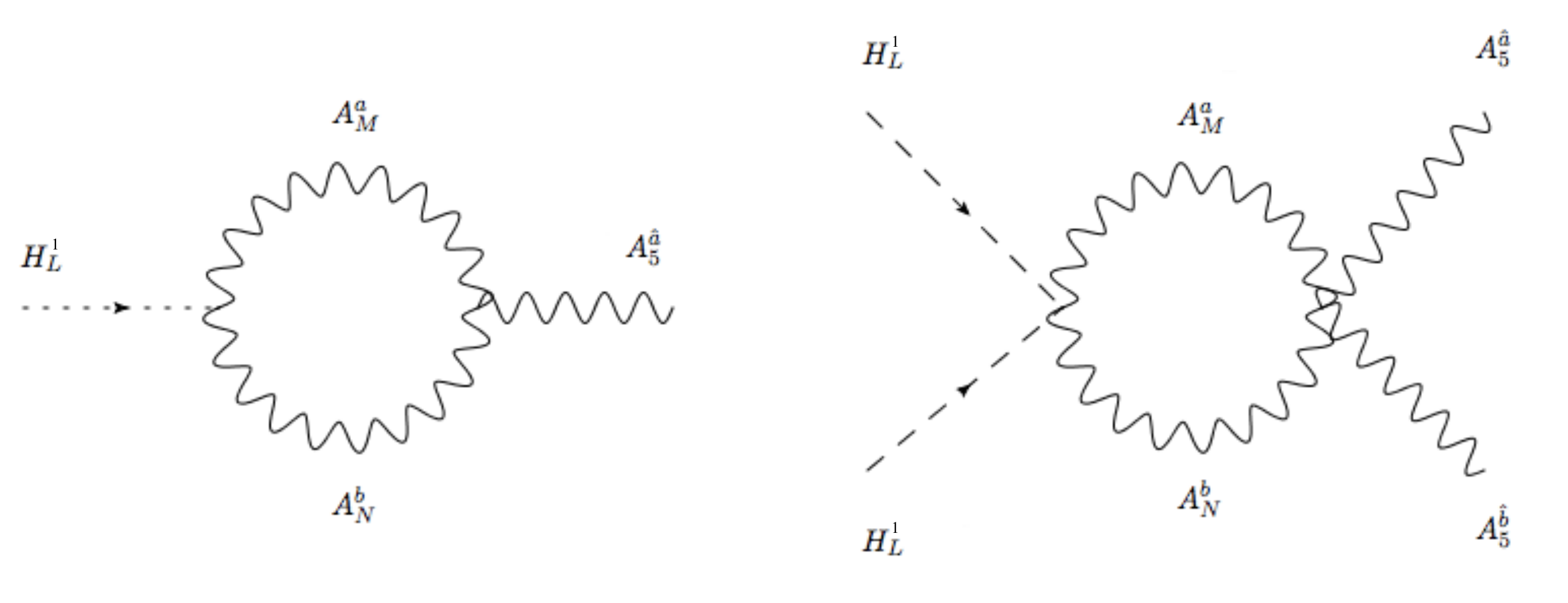} 
\label{fig:fe}
\caption{Feynman diagrams for processes that generate effective interaction terms between scalars after compactification. The vectors in the loop are $A^a_M\in SO(4)$ and the external ones are $A^{\hat{a}}_5\in SO(5)/SO(4)$. After compactification, the former become W and Z bosons and the latter becomes the Hosotani scalar.}
\end{figure}

\begin{eqnarray}
V_{int}&=&\nonumber\Big\{-\mu_{12}^2~Tr[H^\dagger_L H_h] +h.c.
+\lambda_3 ~Tr[H^\dagger_LH_L H^\dagger_h H_h] \Big\}\Big|_H ~,
\end{eqnarray}
where $\{ \}|_H$ means that what is inside must respect H-parity after EWSB. In the model EWSB happens different from the SM, due to the Hosotani mechanism there is a residual symmetry $P_H$. To maintain this symmetry, the terms inside $\{ \}|_H$ can't contain odd power terms for $H(x)$ after EWSM. This notation is needed since the $H(x)$ is H-parity odd, but the VEV $v$ accompanying it, is even.
Due to H-parity, $H(x)$ is stable and a viable candidate for dark matter \cite{Hosotani:2010rg}.

\section{Fermions in the bulk and brane} 

The fermion content of this model resembles the one in the HOOS model \cite{Hosotani:2008tx}.
We have three fermion multiplets in the bulk (eq. \ref{femo}), that due to the chosen boundary conditions, these break into:
\begin{eqnarray}
\Psi_{u,d,e}\sim[\textbf{5},Q_X]\to P_L^{u,d,e}\oplus (u,d,e)_R\sim[\textbf{4},Q_X]\oplus[\textbf{1},Q_X]~,
\end{eqnarray}
where the subindices indicate their chirality and $Q_X$ their respective charges under $U(X)$. Since  $SO(4)\simeq SU(2)_L\times SU(2)_R$ the 4-tuplet left-handed fermion behaves as:
\begin{equation}
P_L^{u,d,e}\sim[\textbf{2},\textbf{2},Q_X]
\end{equation}
under $SU(2)_R\times SU(2)_L\times U(1)_X$.
Inside these bidoublets lie the left doublets of the SM and some exotic ones.
The $(u,d,e)_R$ becomes the right-handed fermions in the SM.

To give mass to the exotic fermions's zero modes, we add more fermions in the brane with the following transformation properties under $SU(2)_R\times SU(2)_L\times U(1)_X$:

\begin{eqnarray}
\chi^1_R\sim[\textbf{1},\textbf{2},7/6],\hspace{5mm}\chi^2_R\sim[\textbf{1},\textbf{2},1/6],\hspace{5mm}\chi^3_R\sim[\textbf{1},\textbf{2},-5/6],\hspace{5mm}\chi^4_R\sim[\textbf{1},\textbf{2},-1/2],
\label{fers}
\end{eqnarray}
where $R$ indicates their chirality. It is shown in \cite{Hosotani:2009qf} that this set of fermions are anomaly free.

Each bidoublet $P_L$ breaks into two doublets after the breaking of $SU(2)_R$. Some of them couple to the brane fermions through the terms (since there is no confusion, we omit chirality subindices):
\begin{equation}
-y_1\bar{\chi}^1 P^u \tilde{H}_R^T-y_2\bar{\chi}^2 P^d \tilde{H}_R^T-y_3\bar{\chi}^3 P^d H_R^\dagger-y_4\bar{\chi}^4 P^e \tilde{H}_R^T+h.c
\end{equation}
that, after breaking right symmetry, give masses to exotic fermions
that are of $O(10\:TeV)$ and decouple from the theory.

The Hosotani scalar obtains a VEV $\sim O(100GeV) $ and it couples to the fermions on the IR brane giving them masses. The fact that the VEV comes from the bulk is important because it will be lowered in the IR brane when the IR fields absorb the warp factor (the same way it is done for the hierarchy problem) \cite{Randall:1999ee}. The brane fermion masses coming from the Hosotani scalar turn out to be $\sim O(10^{-13} GeV)$ and thus completely negligible.

The Hosotani mechanism gives masses to one fermion per multiplet and mixes fermions with equal $U(1)_{EM}$ charge. The fermions that obtain mass this way are identified with the up and down quark and electron. This is calculated in \cite{Hosotani:2008tx} and these fermion masses are proportional to:
\begin{equation}
m_f\propto m_{KK}\sqrt{1-4c_f^2}~\label{hofm},
\end{equation}
where $c_f$ is the fermion's kink mass.

In this model there is another contribution te fermion masses coming from the VEV of $H_L$. This is due to the terms:
\begin{equation}
-\sum_{\psi=u,d,e} y_\psi P_L^\psi H_L \psi_R^\dagger+h.c.
\end{equation}
that con tributes to masses of every fermion with EM charge $Q_{EM}=\{2/3,-1/3,-1\}$.

To account the fermion families we must have three copies of each fermion we have thus far.
Since the exotic fermions are very heavy and decouple from the theory, we have effectively the complete SM fermion content.

\section{Phenomenology}

We choose $e^{kL}=10^{15}$ so that the hierarchy problem is explained by this warp factor.

The main constraints for the extra dimensions parameters, in this model, come from KK mode's contributions to the $S,T,U$ parameters \cite{Carena:2007ua,Beringer:1900zz,Agashe:2003zs}. Since this model has a custodial symmetry, the constraint is considerably lowered to \cite{Carena:2007ua}:
$$m_{KK} \gtrsim 850 GeV.$$
This is a very low energy that would be in conflict with KK gluon constraints. This will be worked out below.

\begin{figure}[!h]
\centering
 \includegraphics[scale=1.3]{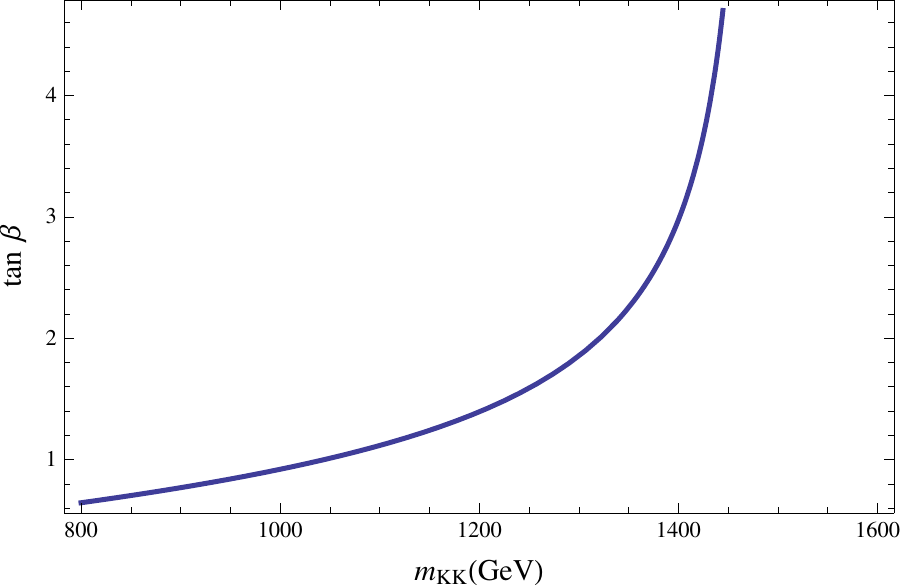} 
\caption{$\tan\beta$ dependence KK mass scale.}
\label{fig:tan}
\end{figure}

The Hosotani scalar VEV depends on $m_{KK}$ and since both scalars give masses to $W$ and $Z$ bosons then $\tan \beta=v/v_L$ depends on it too, as can be seen in figure \ref{fig:tan}.

The left scalar $h(x)$ should behave as the boson recently measured by ATLAS and CMS \cite{Chatrchyan:2013lba, atlasconf}. The couplings of this scalar to fermions are SM like with an extra factor of $\cos\beta$, while its couplings to $W$ and $Z$ bosons have a factor of $1/\cos\beta$. As we see in figure \ref{fig:tan}, $\cos\beta$ decreases with $m_{KK}$. If we want the couplings of the left scalar to be SM like, then $m_{KK}$ should be low. In our calculations we use $m_{KK}=900$ GeV.

The pseudoscalar and charged scalar obtain a mass $m_{A^0,H^\pm}\sim\mu_{12}$ that needs to be $\mu_{12}\gtrsim 100 GeV$ to comply with current constraints \cite{Beringer:1900zz}.
The Hosotani scalar $H(x)$ obtains a mass which depends on $m_{KK}$  and the effective potential \cite{Haba:2008dr, Hosotani:2008tx}. The Hosotani scalar can be light enough for the left scalar to decay into it with width:
\begin{equation}
\Gamma_{h\to HH}=\frac{1}{4\pi m_h}| \lambda_3 v_L  |^2 \sqrt{1-4m_H^2/m_h^2}~. \label{decay}
\end{equation}

\begin{figure}[!h]
\centering
 \includegraphics[scale=1.3]{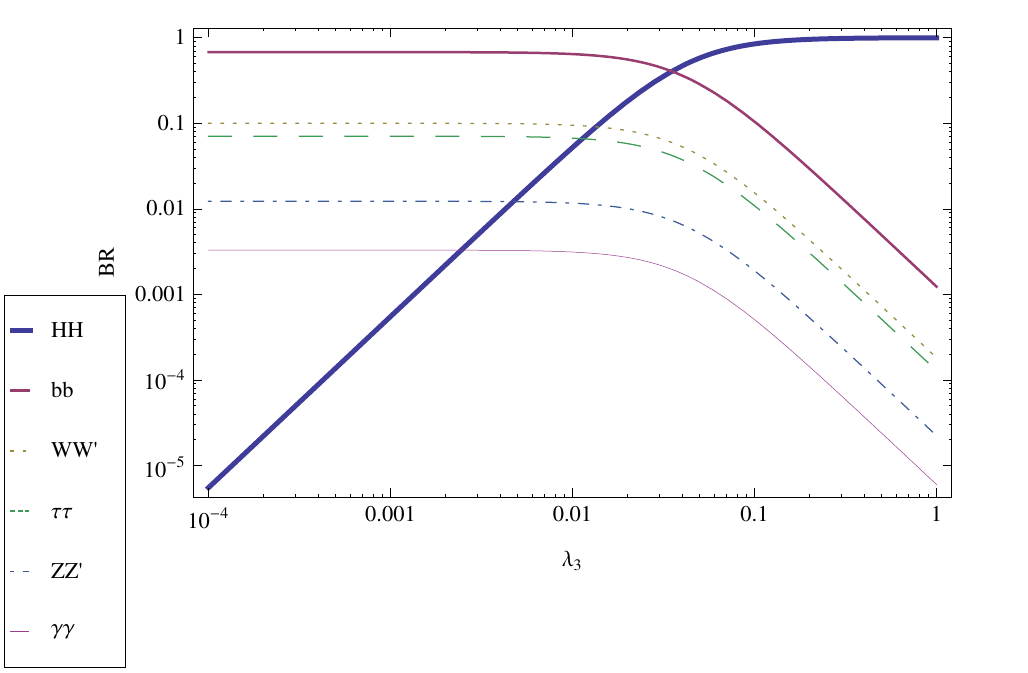} 
\caption{Branching ratios of the left scalar depending on the Hosotani-left scalar couplings.}
\label{fig:dec}
\end{figure}

We anayze the case when the Hosotani scalar's mass is $m_H=60 GeV$, which is a typical mass with the parameters given above. In figure \ref{fig:dec} we plot the $h(x)$ Brs varying the coupling of the scalars $\lambda_3$.

We see in figure \ref{fig:dec} that for a coupling $\lambda_3\sim 1$, the decay into scalars is the main channel. This coupling can be small so that it almost vanishes and the channel is unnoticeable. In that case we obtain the BR for the $W, Z, b$ compared to the SM ones as:
\begin{equation}
\sigma^{Z,W}/\sigma_{SM}^{W,Z}=0.5,~~~ \sigma^{b}/\sigma_{SM}^{b}=1.2, \label{brs}
\end{equation}
consistent with current measurements \cite{Chatrchyan:2013lba, atlasconf}.

Every fermion receives mass contributions from both scalars, depending on its kink mass parameter $c$ and its coupling to the left scalar: 
\begin{equation}
m_f=m_f^h(c_f,m_{KK})+\frac{y_fv_L}{\sqrt{2}},
\end{equation}
where $m_f^h$ is the Hosotani mechanism's contribution to the tau's mass and $y_f$ is it's usual Yukawa coupling. This, in principle, leaves free the fermion - Higgs coupling $y_f$ and can account for the small differences in couplings coupling detected in recent experiments \cite{Chatrchyan:2014nva}.

Since we have an effective THDM we need to make sure that the extra scalars don't generate excessive Flavor Changing Neutral Currents (FCNCs).
The most stringent constraints for FCNCs come from the decay $\mu\to e\gamma$ in the lepton sector and $\bar{K}_0-K_0$ mixing in the quark sector \cite{Beringer:1900zz}. This constraints, together with the measured Higgs to fermion couplings, restricts the masses coming from the Hosotani mechanism to be:
\begin{equation}
m_q^h\lesssim 20 MeV,\hspace{7mm}m_l^h \lesssim 100MeV,\label{fcc}
\end{equation}
where $m_q^h$ is the contribution of the Hosotani mechanism to any quark's mass and the $m_l^h$ is its contribution to any lepton's mass. This constraint fixes the third family Yukawa couplings to be SM-like, agreeing with recent measurements, and leaves some freedom in the first family Yukawa couplings.

The gauge boson's KK modes are another important source of FCNCs. In this model, these are sufficiently lowered by the rs-GIM mechanism \cite{Agashe:2004cp, Huber:2003tu} and with the used value of $m_{KK}$ we can avoid constraints. This is not the case when we take into account KK gluon contributions. Fortunately the constraint in eq. \ref{fcc} restricts the fermion's kink mass through eq. \ref{hofm} to be $c_f\sim 1/2$ for all fermions. This results in a practically flat fifth dimensional profile for all fermions, strongly reducing FCNCs coming from any KK gauge boson.

The decay of $h(x)$ to photons is enhanced by the charged scalar and the different couplings to fermions and $W$ boson, with the setup given above, we obtain a BR compared to the SM:
\begin{equation}
\sigma^{\gamma\gamma}/\sigma_{SM}^{\gamma\gamma}=1.2
\end{equation}
consistent with experimental data \cite{Chatrchyan:2013lba}. 

\subsection{Hosotani scalar as a Dark Matter candidate}

As we have showed above, the geometry of the extra dimension and its compactification give us a residual symmetry, the H parity. The Hosotani scalar is odd under H parity, and everything else is even, thus the Hosotani scalar is completely stable. This is an interesting property that allows it to be a Dark Matter candidate. We now analyze its relic density and its cross section with nucleons.

The interaction of the Hosotani scalar with fermions, at tree level, is only mediated by the left scalar. Thus, its cross section with nucleons and its relic density depend strongly on the Hosotani-left scalar coupling $\lambda_3$.

\begin{figure}[!h]
\centering
 \includegraphics[scale=1]{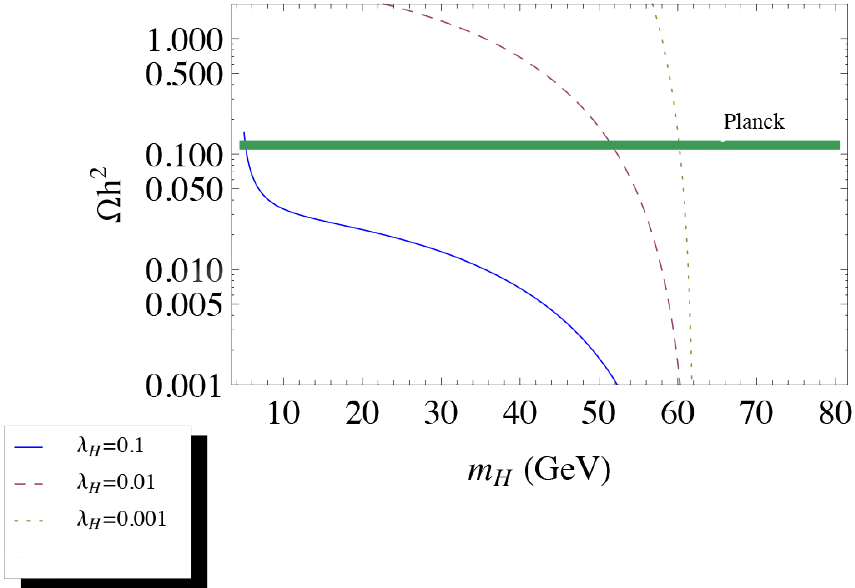} 
\caption{Relic density of the Hosotani scalar depending on its mass and its coupling to the left scalar. Planck measurement shown.}
\label{fig:relic}
\end{figure}

To comply with experimental measurements from Planck, our model should generate a relic abundance for the Hosotani scalar of $\Omega h^2 = 0.1199\pm0.0027$ \cite{Ade:2013zuv}. We have calculated its relic density using the "freeze-out approximation" \cite{Griest:1995gs}, in this model it depends strongly on the Hosotani scalar's mass $m_H$ and its coupling with the left scalar $\lambda_3$. 
In figure \ref{fig:relic} we show the relic density of the Hosotani scalar depending on these parameters.

\begin{figure}[!h]
\centering
 \includegraphics[scale=1]{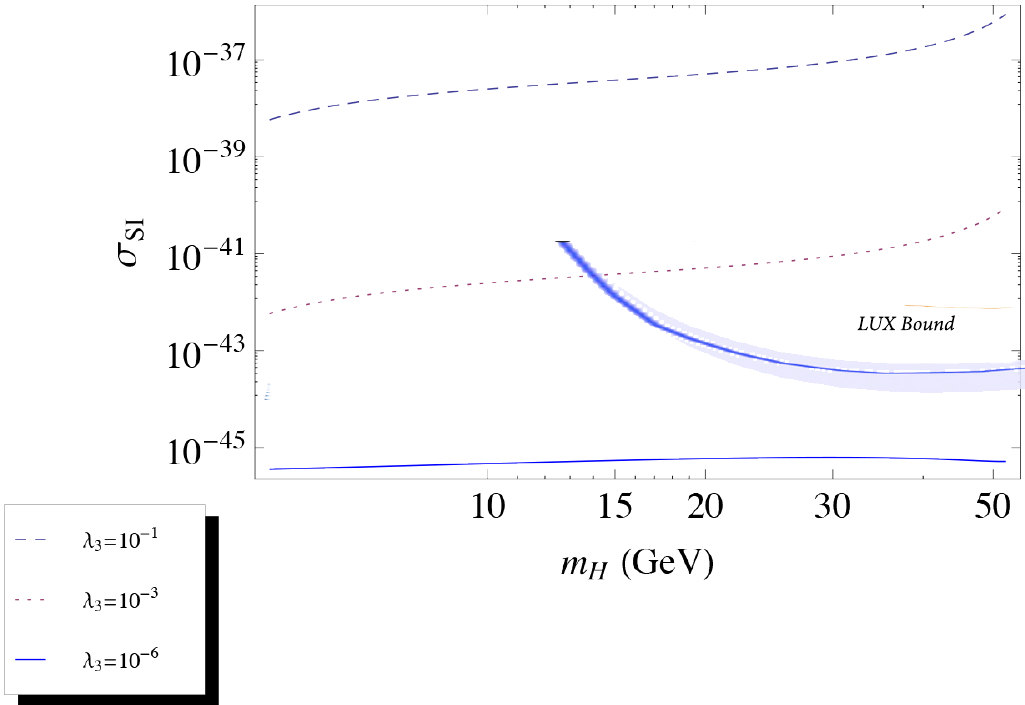} 
\caption{Hosotani scalar's spin averaged nucleon cross section. LUX bound shown. Cross sections bigger than this bound are ruled out.}
\label{fig:css}
\end{figure}

Finally we calculate the spin averaged Hosotani scalar's cross section with nucleons $\sigma_{SI}$. This cross section is shown in figure \ref{fig:css}.

We see from both plots that there are parameter areas that comply with both measurements, making the Hosotani scalar a good dark matter candidate.

\section{Conclusions}

We have been able to build a realistic model that incorporates the idea of Gauge Higgs Unification but not in the usual sense. Gauge Higgs Unification proposes that the Higgs boson comes from the higher dimensional components of a gauge field. This idea needs too many extra fields to work, so in our model it is not the Higgs boson the one that comes from GHU, this boson becomes dark matter.

We have a slight modification of the HOOS model with gauge symmetry based on the group $SO(5)\times U(1)$ and it lives in the rs spacetime. The gauge symmetry is broken to $SU(2)_L\times SU(2)_R\times U(1)_X$ via orbifolding, and we have a custodial symmetry that helps to comply with electroweak precision tests.

After compactification the higher dimensional components of the gauge fields behave as a scalar that obtains a VEV through the Hosotani mechanism. This Hosotani scalar is one of the scalar fields that break the LR symmetry but not the Higgs boson, as in usual GHU. Furthermore, due to symmetries of the model, it is stable and therefore a viable dark matter candidate.

There are two additional scalar fields, charged under $SU(2)_{L,R}$ that live on the IR brane and obtain a VEV, breaking the LR symmetry. The left doublet contains the Higgs boson and we fit its observables to the measured ones. The fact that these scalars live on the IR brane of the rs spacetime solve the hierarchy problem.

The $SU(2)_R\times U(1)_X$ symmetry breaks at the 10 TeV scale leaving effectively, at low energies, a THDM with $SU(2)_L\times U(1)_Y$ gauge symmetry, whose scalar sector contains a charged scalar and pseudoscalar within experimental constraints.

All the fermion content of the SM lives on the bulk. These fermions receive their mass from a combination of both, the Hosotani mechanism and the left scalar VEV. 

The model has FCNCs generated by KK gauge bosons and the effective THDM. These can be made sufficiently small to comply with experimental constraints.

We showed that the Hosotani scalar is completely stable, it has a very small cross section and it can have the necessary relic density to be an interesting Dark Matter candidate.

We have been able to fit all the SM observables, including recent measurements from the Higgs boson. We evade all constraints from extra dimensions, left-right symmetries and extra fermions. The model is well constrained by this, but still viable.

\section{Acknowledgement}

Really special thanks go to Alfredo Aranda for his fundamental help in this work. This work was supported in part by CONACYT.

\bibliographystyle{unsrt}

\end{document}